\documentclass[a4,article,nofootinbib,twocolumn]{revtex4}

\usepackage{graphicx}
\usepackage{dcolumn}
\usepackage{bm}
\usepackage{color}
\usepackage{amsmath}%
\usepackage{amsfonts}%
\usepackage{amssymb}%
\usepackage{breqn}
\usepackage{fixfoot}

\usepackage{bm}
\usepackage{mathrsfs}
\usepackage{amsthm}
\usepackage{breqn}
\usepackage{doi}
\usepackage{float}
\usepackage[normalem]{ulem}

\newcommand{\mc}{\mathcal}
\newcommand{\cp}{\times}

\newcommand{\bol}{\boldsymbol}

\newcommand{\abs}[1]{\left\lvert{#1}\right\rvert}

\newcommand{\lr}[1]{\left({#1}\right)}

\newcommand{\mf}{\mathfrak}
\newcommand{\p}{\partial}

\makeatletter
\let\cat@comma@active\@empty
\makeatother

\begin{document}
\title{The galactic rotation curve of a magnetized plasma cloud}
\author{N. Sato}
\affiliation{Research Institute for Mathematical Sciences, Kyoto University, Kyoto 606-8502, Japan}
\date{\today}

\begin{abstract}
The rotation curve of a magnetized plasma cloud orbiting in the gravitational
potential of a galaxy is calculated by statistical arguments.
The working assumption is that a mild magnetic field, decreasing with distance
from the galactic center, permeates space. 
It is shown that the resulting probability density is compatible with a flat rotation curve.
\end{abstract}


\maketitle

\begin{normalsize}

A galaxy rotation curve is the functional relationship between the velocity with which 
matter rotates around the galactic center, and the radial distance from the galactic center. 
The discrepancy between the rotation curve predicted by exact balance between
gravitational and centripetal accelerations and that obtained from observations \cite{Rubin,Rubin2}
is considered indirect evidence for the existence of dark matter \cite{Sanders} or for the need of modifications of Newtonian gravity \cite{Gaugh}.
Indeed, the galactic mass calculated from observed light distributions \cite{Freeman}
does not match the rotational velocity measured from neutral hydrogen emission in the radio frequency spectrum \cite{Gaugh2}. The dark matter density profile needed to justify 
galactic rotation curves can be modeled with the aid of N-body simulations \cite{Navarro,Read,Ciotti}.

The purpose of this paper is to investigate the effect of a galactic magnetic field
on the rotation curve of a plasma cloud (e.g. an H II region) orbiting in the gravitational potential of a galaxy.
Galaxies generally exhibit magnetic fields of several $\mu\, G$ 
that can be seen from radio synchrotron emission both in central regions, spiral arms, and among spiral arms \cite{Beck2,Chamandy}. Magnetic fields play an important role in galactic dynamics and
evolution \cite{Beck,Birnboim}, and their effect has been studied with MHD simulations \cite{Wang}. 
The reason for exploring this problem is that, even in a mildly ionized
system (such as the hydrogen clouds \cite{Ferriere} inside and at the edge of galaxies), any rotational momentum
acquired by the ionized component due to magnetization will be gradually transferred to the neutral component by thermalization, thus providing an effective mechanism to stir the whole gas.

In order to determine the rotation curve of the plasma,
we will apply the arguments of equilibrium statistical mechanics.
This has to be done with care, because the time needed to reach the maximum entropy state
for an ideal collisionless system evolving under the effect of long-range interactions \cite{Chavanis2} such as gravity \cite{LyndenBell,Chavanis} diverges with the number of particles in the ensemble \cite{Teles}.
Instead, quasistationary states depending on the initial conditions of the system can be achieved
which exhibit heterogeneous structures \cite{Pakter,Antoniazzi}.   

In our model, the quasistationary states are obtained as local entropy maxima
that entail information on the initial configuration due to the presence of topological constraints
that cannot be violated over the timescale of the quasirelaxation (hence the constraints behave 
as adiabatic invariants \cite{Yoshida}).
This procedure has a concrete justification once ergodicity is not imposed on the whole phase space,
but only over the regions of phase space (the level sets of the constraints) that are dynamically accessible.
Such regions are mathematically determined by the Casimir invariants spanning the null-space of the Poisson operator that characterizes the Hamiltonian formulation \cite{Morrison} of single particle dynamics \cite{Sato,Sato3}.

First, let us determine the rotation curve of a neutral gas within a potential $\phi=\phi\lr{r,z,\theta}$, where $\lr{r,z,\theta}$ is a cylindrical coordinate system. This will be useful when comparing the result for a magnetized plasma system. 
The assumption on $\phi$ is that 
$\lim_{R\rightarrow\infty}\phi=0$, with 
$R=\sqrt{r^2+z^2}$ the radius of a spherical coordinate system.  

To formulate a statistical model of the gas, we need to identify the energy $H$ (Hamiltonian) of
single particle motion (note that, in principle, the particle may represent any mass distribution that is
negligible in size with respect to the characteristic scales of the system), any constant of motion $C$, and most importantly the invariant measure $dV_{\rm{I}}$ (if it exists) associated to the equations of motion.
The energy of a particle of mass $m$ is:
\begin{equation}
H=\frac{1}{2}m v^2+\phi=\frac{1}{2}m\lr{v^{2}_{r}+\omega^2 r^{2}+v^{2}_{z}}+\phi,
\end{equation}
where $v_{r}=\dot{r}$, $\omega=\dot{\theta}$, and $v_{z}=\dot{z}$.
Setting $\bol{z}=\lr{v_{r},r,\omega,\theta,v_{z},z}$, the equations of motion can be cast in the form
$\dot{\bol{z}}=\mc{J}\p_{\bol{z}}H$, with $\mc{J}$ the antisymmetric matrix:
\begin{equation}
\frac{1}{m}\begin{bmatrix} 0&-1&\frac{2\omega}{r}&0&0&0\\ 
                           1&0&0&0&0&0\\ 
													 -\frac{2\omega}{r}&0&0&-\frac{1}{r^2}&0&0\\ 
													 0&0& \frac{1}{r^2}&0&0&0\\
													 0&0&0&0&0&-1\\
													 0&0&0&0&1&0 \end{bmatrix}.
\end{equation} 
Since $\mc{J}$ satisfies the Jacobi identity $\{f,\{g,h\}\}+\{g,\{h,f\}\}+\{h,\{f,g\}\}=0$ with $\{f,g\}=\p_{z^{i}} f\mc{J}^{ij}\p_{z^{j}}g$, where $f$, $g$, and $h$ are smooth functions, the equations of motion for $\bol{z}$ are a Hamiltonian system.
Then, the invariant measure is provided by the phase space volume element:
\begin{equation}
dV_{\rm{I}}=dp_{x}\,dx\,dp_{y}\,dy\,dp_{z}\,dz=m^{3}r^2\, dv_{r}\,dr\, dv_{z}\,dz\, d\omega\, d\theta.\label{dVIgas}  
\end{equation}
Here, the $p_{i}$s are the standard canonical momenta.
In the presence of a symmetry in the Hamiltonian, by Noether's theorem, there is
an associated constant of motion. For the case of interest, we consider the possibility that
the potential $\phi$ is independent of the angle $\theta$ (as is the case of Keplerian rotation).
The relevant invariant is therefore the $z$-component of the angular momentum, $L_{z}=\omega r^2$. 
If one drops (reduces) the phase $\theta$, the system becomes $5$-dimensional, and the Poisson matrix $\mc{J}$
becomes degenerate. The null-space will be spanned by the gradient of $L_{z}$, implying that
any choice of $H$ will preserve the vertical component of the angular momentum ($\dot{L}_{z}=\p_{\bol{z}}L_{z}\cdot\mc{J}\p_{\bol{z}}H=0$ if $\mc{J}\p_{\bol{z}}L_{z}=\bol{0}$), which becomes a Casimir invariant.

The entropy $S$ of the gas is given by the functional:
\begin{equation}
S=-\int{f\log{f}}\,dV_{\rm{I}},
\end{equation}
where integration is performed on $\mathbb{R}^6$, and $f$ is the probability density
on the invariant measure $dV_{I}$ (note that it is crucial that $f$ is defined on the measure \eqref{dVIgas} because the argument of the logarithm must be a probability, see \cite{Sato}).
The equilibrium probability density can now be evaluated by maximization of entropy under the constraints of total energy $E=\int{fH} \,dV_{\rm{I}}$, total particle number $N=\int{f}\, dV_{\rm{I}}$, and total vertical angular momentum $\Omega_{z}=\int{fL_{z}^2}\,dV_{\rm{I}}$, according to the variational principle $\delta\lr{S-\alpha N-\beta E-\gamma \Omega_{z}}=0$ ($\alpha$, $\beta$, and $\gamma$ are real constants). 
The functional form of the integrand in $\Omega_{z}$ is determined by observing that the conserved
quantity $L_{z}$ must enter the constraint $\Omega_{z}$ in the same way it enters the Hamiltonian $H$,
because, once multiplied by the appropriate chemical potential, $\Omega_{z}$ must represent
the change in energy due to addition or removal of the constant of motion. 
The result of the variation is:
\begin{equation}
f=\frac{1}{Z}\exp\left\{-\beta H-\gamma L^2_{z}\right\},
\end{equation} 
with $Z=e^{1+\alpha}$ a normalization constant such that $\int{f}\,dV_{\rm{I}}=1$.
The corresponding spatial density $\rho$ is:
\begin{equation}
\rho=\int f d^3\bol{p}=m^{3}r\int fdv_{r}dv_{z}d\omega=\frac{e^{-\beta\phi}}{Z}\frac{\lr{2\pi m/\beta}^{3/2}}{\sqrt{1+\frac{2\gamma }{\beta m}r^2}}.\label{rhoz}
\end{equation}
Similarly, denoting $v_{\theta}=\omega r$, the (self) rotation curve is:
\begin{equation}
\langle v^{2}_{\theta}\rangle=\frac{\int{ f v^{2}_{\theta}}\,d^{3}\bol{p}}{\int{ f }\,d^{3}\bol{p}}=\frac{(\beta m)^{-1}}{1+\frac{2\gamma}{\beta m}r^2}.\label{vdz}
\end{equation}
If the system is symmetric with respect to all rotations, we may replace $\Omega_{z}$ with the total angular momentum $\Omega=\int{f L^2}\,dV_{\rm{I}}$ and obtain analogous formulas:
\begin{subequations}\label{L2}
\begin{align}
f&=\frac{1}{Z}\exp\left\{-\beta H-\gamma L^2\right\},\\
\rho&=\frac{e^{-\beta\phi}}{Z}\frac{\lr{2\pi m/\beta}^{3/2}}{{1+\frac{2\gamma }{\beta m}R^2}},\\
\langle v^{2}_{\rm{rot}}\rangle&=\frac{2(\beta m)^{-1}}{1+\frac{2\gamma}{\beta m}R^2},
\end{align}
\end{subequations}
 where $v^{2}_{\rm{rot}}=v^2-\dot{R}^2$ is the rotational velocity.

Now consider a body, say a star, orbiting within the gas cloud at radius $R$.
The rotation curve $v_{\ast}\lr{R}$ of the body can be estimated by assuming that
it does not participate to the gas relaxation process.
In such case the centripetal acceleration $v^{2}_{\ast}/R$ and the gravitational acceleration $GM\lr{R}/R^2$,
with $G$ the gravitational constant and $M\lr{R}$ the gas cloud mass within $R$, must balance.
In the scenario of equation \eqref{rhoz} with $\gamma\neq 0$, we have $\rho\sim R^{-1}$ at large radii on the plane $z=0$. On the other hand, from the balance equation $GdM=d\lr{Rv^{2}_{\ast}}$:
\begin{equation}
\rho=\frac{v_{\ast}^{ 2}}{4\pi G R^2}\lr{1+2\frac{R}{v_{\ast}}\frac{dv_{\ast}}{dR}},\label{Kep}
\end{equation} 
which gives $v_{\ast}\sim \sqrt{R}$. When the conservation of vertical angular momentum is broken,
i.e. $\gamma=0$, one has $\rho\sim \rm{const.}$, leading to $v_{\ast}\sim R$.
Similarly, in the scenario of equation \eqref{L2}, $\gamma\neq 0$ implies $\rho\sim R^{-2}$ and $v_{\ast}\sim \rm{const.}$, and $\gamma =0$ implies $\rho\sim\rm{const.}$ and $v_{\ast}\sim R$.
Nevertheless, from equations \eqref{rhoz}, \eqref{vdz}, and \eqref{L2} it is clear that
the constraint of total angular momentum is not compatible with a non-decreasing rotation curve
for decreasing density. The scenario is different if, in addition to the potential $\phi$, we
assume that a magnetic field $\bol{B}$ permeates the gas cloud, and that particles can interact electromagnetically, as in the case of an ionized gas.

An ionized gas is a plasma. In a magnetic field, a charged particle 
performs a periodic motion around the local magnetic field line, the cyclotron gyration.
The angular frequency of the gyration is given by $\omega_{c}=eB/m$, with $e$ the electric charge and $B=\abs{\bol{B}}$ the magnetic field strength.
For a proton in a typical galactic environment at the hydrogen ionization temperature $T\sim 10^4\,K$ within a magnetic field $B\sim~1\,nT$, $\omega_{c}\sim 0.1\,Hz$. The radius of the gyration is given by $r_{c}=mv_{c}/eB\sim 10^5\,m$, where $v_{c}$ is the velocity of the gyration. 
The adiabatic invariant associated to cyclotron motion is the magnetic momentum $\mu=m v^{2}_{c}/2B$.
The motion of a magnetized particle can be expressed in terms of the averaged dynamics of the guiding center (the point in the middle of the cyclotron orbit). The guiding center Hamiltonian is:
\begin{equation}
H_{g}=\mu B+\frac{m}{2}v^{2}_{\parallel}+\phi,\label{Hgc}
\end{equation}
with $v_{\parallel}$ the velocity along the magnetic field.
In the following we assume that the magnetic field has the expression $\bol{B}=\nabla\psi\cp\nabla\theta$,
with $\psi=\psi\lr{r,z}$ the so called flux function. In the case of a point dipole magnetic field $\psi=\mc{M} r^{2}/\lr{r^2+z^2}^{3/2}$ with $\mc{M}$ a physical constant.
It is known \cite{Yoshida} that the drift velocity $v_{\theta}$ around the $z$-axis (such as that of a gravitational orbit in the $z=0$ plane) does not contribute to the guiding center Hamiltonian \eqref{Hgc}.
This is because the canonical angular momentum $p_{\theta}=m r v_{\theta}+e\psi$ satisfies $p_{\theta}\approx e\psi$ due to the smallness of the particle mass, and therefore the kinetic term $\lr{p_{\theta}-e\psi}^{2}/2 m r^2$ is negligible. For a particle moving with velocity $v_{\theta}\sim 10^5\, m/s$ in a dipole magnetic field $B\sim 1\,nT$ with characteristic length scale of $30\,kpc$ one has $mrv_{\theta}/e\psi\sim m e^{-1} \, 10^{-7}\,C/kg$. Hence, this approximation remains true even for more massive objects bearing a small charge and moving in a galactic environment. 
Setting $\bol{z}=\lr{v_{\parallel},\ell,\psi,\theta,\mu}$, with $\ell$ a length coordinate along the magnetic field, the equations of motion for the guiding center (see \cite{Sato2}) can be cast in the form $\dot{\bol{z}}=\mc{J}\p_{\bol{z}}H$, where $\mc{J}$ is the antisymmetric matrix:
\begin{equation}
\frac{1}{e}\begin{bmatrix} 0&-em^{-1}&0&v_{\parallel}q_{\ell}&0\\ 
                           em^{-1}&0&0&-q&0\\ 
													 0&0&0&-1&0\\ 
													 -v_{\parallel}q_{\ell}&q&1&0&0\\
													 0&0&0&0&0\\
		       \end{bmatrix}.
\end{equation} 
Here $q=\nabla\psi\cdot\nabla\ell/\abs{\nabla\psi}^2$ is
a geometrical quantity measuring the non-orthogonality of the coordinate system $\lr{\ell,\psi,\theta}$,
and $q_{\ell}$ its $\ell$-derivative. The degrees of freedom are only $5$ because the phase $\theta_{g}$ of the cyclotron gyration does not bear physical significance at the length and time scales of interest.  
Since $\mc{J}$ satisfies the Jacobi identity, this system is again Hamiltonian. However, the rank of the matrix is $4<5$, implying that there is a null-space of dimension $1$. It is immediate to verify that the vector $\p_{\bol{z}}\mu$ spans the null-space of $\mc{J}$,
making $\mu$ a constant of motion since $\dot{\mu}=\dot{\bol{z}}\cdot\p_{\bol{z}}\mu$. Integrals like $\mu$ depending only on the form of the Poisson matrix $\mc{J}$, which entails the geometric properties of space, are Casimir invariants. This is in contrast with conservation of angular momentum, which is a consequence of a symmetry in the Hamiltonian, i.e. a property of matter (remember that conservation of angular momentum can be transformed to a property of space by reducing the associated phase variable).
Due to their `spatial' nature, Casimir invariants can distort the metric of the effective phase space of the system. Indeed, the invariant measure of guiding center dynamics is:
\begin{equation}
dV_{g}=d\mu\,dv_{\parallel}\,d\ell\,d\psi\, d\theta=B\,d\mu\, dv_{\parallel}\, dx\, dy\, dz.
\end{equation} 
Observe how the natural coordinate metric of the system is not $dx\, dy\, dz$, but the measure $d\ell\, d\psi\,d\theta=B\,dx\, dy\, dz$ weighted by magnetic field strength.
Hence, while a magnetic field cannot perform work on a particle, it can `bend' space.
This has important consequences for statistical mechanics,
since the proper entropy measure must now have the form (see \cite{Sato}):
\begin{equation}
\Sigma=-\int{P\log{P}}\,dV_{g},
\end{equation}
 where $P\lr{\bol{z}}$ is the probability density on $dV_{g}$.
Maximization of entropy under the constraints of total energy $E_{g}=\int{PH_{g}}\,dV_{g}$,
total particle number $N_{g}=\int{P}\,dV_{g}$, and total magnetic momentum $\mf{M}=\int{P\mu}\,dV_{g}$
gives the probability density function:
\begin{equation}
P=\frac{1}{Z_{g}}\exp\left\{-\beta_{g} H_{g}-\gamma_{g}\mu\right\},\label{P}
\end{equation}
with $Z_{g}$ a normalization constant, and $\beta_{g}$ and $\gamma_{g}$ real constants.
The spatial density profile is:
\begin{equation}
\rho_{g}=B\int{P}\,d\mu\,dv_{\parallel}=\left(\frac{2\pi}{\beta m}\right)^{1/2}\frac{e^{-\beta_{g}\phi}}{Z_{g}}\frac{B}{\gamma_{g}+\beta_{g}B}.\label{rhog}
\end{equation}
This density profile was first obtained in \cite{Yoshida}.
Furthermore, it has been shown that the probability density \eqref{P} is a 
maximum entropy stationary solution of the Fokker-Planck equation satisfied by an ensemble of 
magnetized particles (see \cite{Sato,Sato3}).

In order to calculate the rotation curve, first observe that the guiding center equations of motion demand that:
\begin{equation}
\omega=\frac{1}{e}\left[\mu B_{\psi}+\phi_{\psi}+q\lr{\mu B_{\ell}+\phi_{\ell}}\right]-\frac{m}{e}v^{2}_{\parallel}q_{\ell}.
\end{equation}
Here, the subscripts mean derivation.
To simplify the expressions, we shall neglect all factors involving the
geometric term $q$, which is relatively small and exactly vanishes on the plane $z=0$ in the case of a point dipole magnetic field (if needed, the following calculations can be done by keeping $q$).
The velocity of the rotation around the $z$-axis is thus:
\begin{equation}
v_{\theta}=\frac{r}{e}\lr{\mu B_{\psi}+\phi_{\psi}}.
\end{equation}
It follows that:
\begin{dmath}
\langle v^{2}_{\theta}\rangle_{g}=\frac{\int{Pv^{2}_{\theta}}\,d\mu dv_{\parallel}}{\int{P}\,d\mu dv_{\parallel}}=
\frac{r^{2}}{e^2}\left[\frac{2B^{2}_{\psi}}{\lr{\gamma_{g}+\beta_{g}B}^2}+\frac{2B_{\psi}\phi_{\psi}}{\gamma_{g}+\beta_{g}B}+\phi_{\psi}^{2}\right].
\end{dmath}
Let us now compare the behaviors of $\rho_{g}$ and $v_{g}=\langle v^{2}_{\theta}\rangle^{1/2}_{g}$ for different
magnetic fields at large radii $r>>1$ in the plane $z=0$.
We consider magnetic fields such that $B_{\theta}=0$, $B\lr{r,z}=B\lr{r,-z}$, and with asymptotic behavior $B\lr{r,0}=\mc{M}r^{-\lambda}$ for some non-negative real constant $\lambda$. 
For a point dipole magnetic field $\lambda=3$ and $B\lr{r,0}=\mc{M}r^{-3}$. 
We further assume that the potential $\phi$ is dominated by the gravitational pull of a mass $M_{0}$
located in proximity of the center of the system, $R\rightarrow 0$. This mass could represent 
the massive central bulge of a galaxy. The potential has thus the form $\phi=-GM_{0}m/R$.
Notice that, in addition to the self-induced gravitational potential, 
the electric component of the potential energy is neglected under the assumption that, on average, 
the plasma charge is canceled by that of a second plasma with opposite charge.
It follows that $\rho_{g}\lr{r,0}\propto \mc{M}/\lr{\beta_{g}\mc{M}+\gamma_{g}r^{\lambda}}$ 
because $\lim_{r\rightarrow\infty}\phi=0$ and the exponential term in \eqref{rhog} approaches $1$ at large $r$. 
Furthermore, 
\begin{equation}
\begin{split}
B_{\psi}&\lr{r,0}=\lr{B_{r}r_{\psi}+B_{z}z_{\psi}}_{z=0}=\\&\lr{B_{r}\nabla r\cdot\frac{\nabla\theta\cp\nabla\ell}{B}}_{z=0}=-\lr{B_{r}\ell_{z}/rB}_{z=0}=\lambda r^{-2}.
\end{split}
\end{equation}
Here we used $B_{r}\lr{r,0}=-\lambda\mc{M}r^{-\lambda-1}$, $z_{\psi}=0$, and $\ell_{z}=1$ at $z=0$. $\phi_{\psi}\lr{r,0}$ can be evaluated in a similar way:
\begin{equation}
\begin{split}
\phi_{\psi}\lr{r,0}=
-\lr{\frac{\phi_{r}\ell_{z}}{rB}}_{z=0}=
-\frac{GM_{0}m}{\mc{M}}r^{\lambda-3}.\label{phipsi}
\end{split}
\end{equation}
Then, at $r>>1$, we obtain:
\begin{dmath}
v_{g}^2\lr{r,0}=\frac{1}{e^2}\left[\frac{2\lambda^{2}r^{2\lr{\lambda-1}}}{\lr{\gamma_{g}r^\lambda+\beta_{g}\mc{M}}^2}-\frac{2\lambda GM_{0}m r^{2\lambda-3}}{\mc{M}\lr{\gamma_{g}r^\lambda+\beta_{g}\mc{M}}}+\frac{G^2 M_{0}^2 m^2 r^{2\lr{\lambda-2}}}{\mc{M}^2}\right]\sim \lr{\frac{GM_{0}m}{e\mc{M}}r^{\lambda-2}}^2.\label{vB} 
\end{dmath} 
When $\lambda=3$ (point dipole case), $v_{g}\sim {GM_{0}m r}/{e\mc{M}}$. 
Hence, in the presence of a magnetic field scaling as $r^{-3}$, the 
relaxed density profile decreases steeply at large radii, but the rotational velocity exhibits
an opposite increasing behavior. This holds true also for $\lambda > 3$ with $v_{g}$ scaling as $r^{\lambda-2}$.
Furthermore, observe that for $\lambda>3$ the total mass $M\lr{R}$ of the plasma
does not diverge with growing $R$, but converges to a constant value. For example, if $\lambda=6$, 
the spherical approximation $\rho_{g}\propto 1/\lr{\beta_{g}\mc{M}/\gamma_{g}+R^{6}}$ leads to $M\lr{R}\propto\arctan{\lr{R^{3}\sqrt{\gamma_{g}/\beta_{g}\mc{M}}}}\rightarrow \pi/2$ in the limit of large $R$.
When $\lambda=2$ the rotational velocity approaches the constant value $v_{g}\sim {GM_{0}m}/{e\mc{M}}$ at large radii. Therefore, a magnetic field scaling as $r^{-2}$ is compatible with a decreasing density profile and a flat
rotation curve.
For a proton, $v_{g}\sim 1.4~10^{12}\lr{M_{0}/M_{\odot}}/\mc{M}$, with $M_{\odot}$ the solar mass, and $\mc{M}$ having units of magnetic flux $T\,m^2$ and representing the order of the magnetic flux through the galaxy surface. If $M_{0}/M_{\odot}\sim 10^{10}$ and $\mc{M}\sim 10^{18}\, T\,m^{2}$, $v_{g}\sim 10^4\, m/s$. When $0\leq\lambda<2$, the rotation curve falls toward zero at large radii. 
Figure \ref{fig1} shows the radial profiles of $\rho_{g}\lr{r,0}$, $v_{g}\lr{r,0}$, and $\phi\lr{r,0}$ for the case $\lambda=2$. 
We remark that the rotation curve \eqref{vB} breaks down as soon as the particles cease
to be magnetized. Therefore, at distances where the magnetic field becomes
negligible we expect the rotation curve to return to that of a neutral gas, equations \eqref{vdz} and \eqref{L2}.

\begin{figure}[h]
\hspace*{-0.1cm}\centering
\includegraphics[scale=0.15]{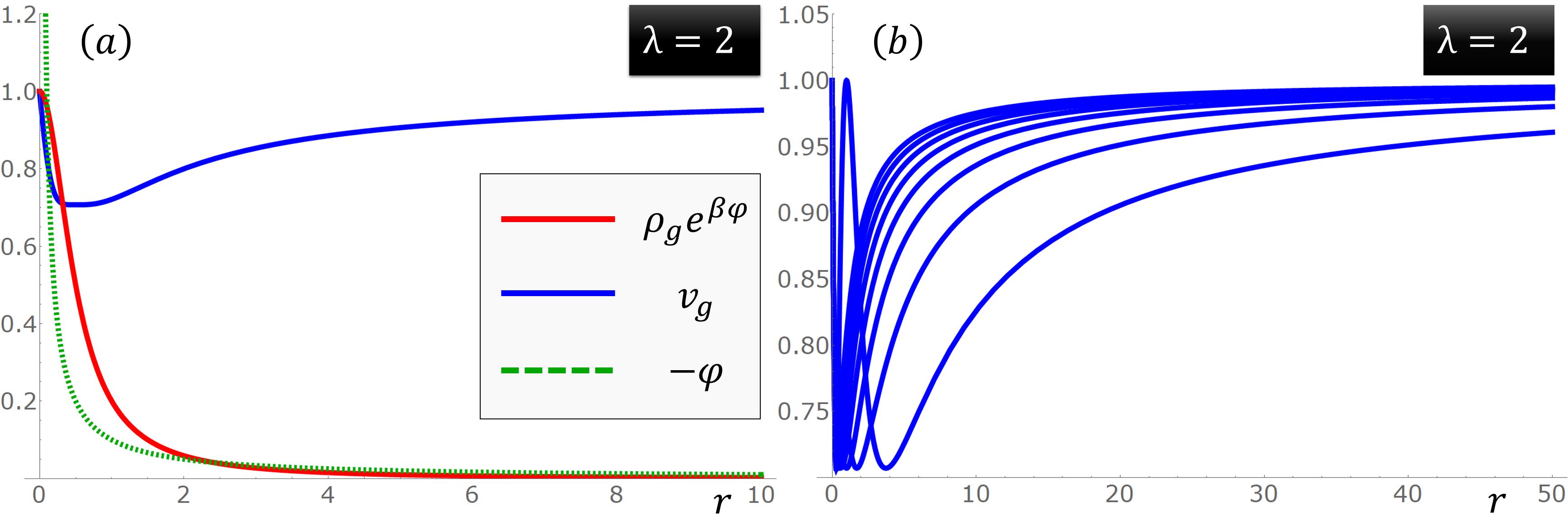}
\caption{\footnotesize (a) Plots of $\rho_{g}\lr{r,0}/e^{-\beta\phi\lr{r,0}}$ (red line), $v_{g}\lr{r,0}$ (blue line), and $-\phi\lr{r,0}$ (dashed green line) for $\lambda=2$. (b) Plots of $v_{g}\lr{r,0}$ for $\lambda=2$ and $\gamma_{g}=n\gamma_{g0}$, with $n=1,...,8$. For higher values of $n$ the asymptotic velocity is achieved at smaller radii. Arbitrary units are used.}
\label{fig1}
\end{figure}	

Finally, consider again a body, say a neutral hydrogen atom or a star, orbiting within the
plasma cloud at a distance $R$. 
Under the assumption that the body does not participate to the relaxation process 
(hence centripetal and gravitational accelerations balance each other), 
we can calculate its rotation curve $v_{\ast}\lr{R}$ from equation \eqref{Kep} by
taking into account of both $M_{0}$ and the plasma mass $M\lr{R}$.
In the approximation of spherical symmetry, $\rho\propto R^{-\lambda}$,
one obtains the asymptotic behaviors $v_{\ast}\sim \sqrt{\log{R}/R}$ for $\lambda=3$,
and $v_{\ast}\sim {\rm{const.}}$ for $\lambda=2$.

The research of N. S. was supported by JSPS KAKENHI Grant No. 18J01729.

\end{normalsize}


\begin{thebibliography}{99}

\bibitem{Rubin} V. C. Rubin, W. K. Ford, and N. Thonnard, ApJ \textbf{225} (1978) pp. L107-L111.
\bibitem{Rubin2} V. C. Rubin, W. K. Ford, and N. Thonnard, ApJ \textbf{238} (1980) pp. 471-487.
\bibitem{Sanders} R. H. Sanders, in \textit{The Dark Matter Problem: A Historical Perspective} 
(Cambridge University Press, Cambridge, 2010), pp. 38-56.
\bibitem{Gaugh} S. S. McGaugh, ApJ \textbf{609} (2004) pp. 652-666.
\bibitem{Freeman} K. C. Freeman, ApJ \textbf{160} (1970) pp. 811-830.
\bibitem{Gaugh2} S. S. McGaugh, Galaxies 2 (2014) pp. 601-622.
\bibitem{Navarro} J. F. Navarro, ApJ \textbf{462} (1996) pp. 563-575.
\bibitem{Read} J. I. Read, J. Phys. G: Nucl. Part. Phys. \textbf{41} 063101 (2014). 
\bibitem{Ciotti} L. Ciotti, ApJ \textbf{471} (1996) pp. 68-81.
\bibitem{Beck2} R. Beck and P. Hoernes, Nature \textbf{379} (1996) pp. 47-49.
\bibitem{Chamandy} L. Chamandy, A. Shukurov, and K. Subramanian, Mon. Not. R. Astron. Soc. \textbf{446} (2015) pp. L6-L10.
\bibitem{Beck} R. Beck, 
Astron. Astrophys. Rev. \textbf{4} 4 (2016).
\bibitem{Birnboim} Y. Birnboim, S. Balberg, and R. Teyssier, Mon. Not. R. Astron. Soc. 
\textbf{447} (2015) pp. 3678-3692.
\bibitem{Wang} P. Wang and T. Abel, ApJ \textbf{696} 1 (2009). 
\bibitem{Ferriere} K. M. Ferri\`ere, Rev. Mod. Phys. \textbf{73} (2001) pp. 1031-1061.
\bibitem{Chavanis2} P. H. Chavanis, AIP Conf. Proc. \textbf{970} 39 (2008).
\bibitem{Chavanis} P. H. Chavanis, J. Sommeria, and R. Robert, ApJ \textbf{471} 385 (1996).
\bibitem{LyndenBell} D. Lynden-Bell and R. Wood, Mon. Not. R. Astron. Soc. \textbf{138}, 495 (1968).
\bibitem{Teles} T. N. Teles, Y. Levin, R. Pakter, and F. B. Rizzato, J. Stat. Mech. P05007 (2010).
\bibitem{Pakter} R. Pakter and Y. Levin, Phys. Rev. Lett. \textbf{106}, 200603 (2011).
\bibitem{Antoniazzi} A. Antoniazzi, D. Fanelli, S. Ruffo, and Y. Y. Yamaguchi, Phys.
Rev. Lett. \textbf{99} 040601 (2007).
\bibitem{Yoshida} Z. Yoshida and S. M. Mahajan, Prog. Theor. Exp. Phys. \textbf{2014} 073J01 (2014).
\bibitem{Morrison} P. J. Morrison, Rev. Mod. Phys. \textbf{70} (1998) pp. 467-521.
\bibitem{Sato} N. Sato and Z. Yoshida, Phys. Rev. E \textbf{93} 062140 (2016).
\bibitem{Sato3} N. Sato and Z. Yoshida, Phys. Rev. E \textbf{97} 022145 (2018).






\bibitem{Sato2} N. Sato, Z. Yoshida, and Y. Kawazura, Plasma Fusion Res. \textbf{11} 2401009 (2016).


\end{thebibliography}
\end{document}